\documentclass[12pt]{iopart}
\usepackage{epsfig}

\def\beq{\begin{equation}}
\def\eeq{\end{equation}}

\def\barr{\begin{eqnarray}}
\def\earr{\end{eqnarray}}
\def\lsim{\raise0.3ex\hbox{$\;<$\kern-0.75em\raise-1.1ex\hbox{$\sim\;$}}}
\def\gsim{\raise0.3ex\hbox{$\;>$\kern-0.75em\raise-1.1ex\hbox{$\sim\;$}}}
\def\bmat{\left( \begin{array}}
\def\emat{\end{array} \right)}

\def\dmsq{\Delta m^2}

\def\nue{{\nu_e}}
\def\nuebar{{\bar{\nu}_e}}
\def\nux{{\nu_x}}
\def\nuxbar{{\bar{\nu}_x}}
\def\ebar{{\bar{e}}}
\def\xbar{{\bar{x}}}

\def\dmbarm{{\overline{\dmsq_m}}}
\def\dmbarc{{\overline{\dmsq_c}}}
\def\tm{{\bar{\theta}_m}}
\def\tc{{\bar{\theta}_c}}

\begin{document}

\title[Signatures of supernova neutrino oscillations]
{Signatures of supernova neutrino oscillations in the Earth 
mantle and core}

\author{A.~S.~Dighe$^1$, M.~Kachelrie\ss$^2$, G.~G.~Raffelt$^2$, and
  R.~Tom\`as$^2$}

\address{$^1$~Tata Institute of Fundamental Research\\
Homi Bhabha Road, Mumbai 400005, India}

\address{$^2$~Max-Planck-Institut f\"ur Physik
(Werner-Heisenberg-Institut)\\
F\"ohringer Ring 6, 80805 M\"unchen, Germany}

\begin{abstract}
The Earth matter effects on supernova (SN) neutrinos can be identified
at a single detector through peaks in the Fourier transform of their
``inverse energy'' spectrum.  The positions of these peaks are
independent of the SN models and therefore the peaks can be used as a
robust signature of the Earth matter effects, which in turn can
distinguish between different neutrino mixing scenarios.  Whereas only
one genuine peak is observable when the neutrinos traverse only the
Earth mantle, traversing also the core gives rise to multiple peaks.
We calculate the strengths and positions of these peaks analytically
and explore their features at a large scintillation detector as well
as at a megaton water Cherenkov detector through Monte Carlo
simulations.  We propose a simple algorithm to identify the peaks in
the actual data and quantify the chances of a peak identification as a
function of the location of the SN in the sky.
\end{abstract}

\section{Introduction}
\label{intro}

The neutrino spectra that arrive at the Earth from a core collapse
supernova (SN) have information about the neutrino masses and mixings
encoded in them. The $\sim 20$ neutrinos observed from SN~1987A were
extensively used trying to constrain the solar neutrino parameters as
well as $\theta_{13}$ and the neutrino mass
hierarchy~\cite{SN87a,SN87b,SN87c,SN87d}.  The determination of
neutrino parameters depends crucially on our understanding of the
primary neutrino fluxes produced inside the SN.  In spite of large
uncertainties on these primary fluxes, some of their robust features
may be exploited to identify the type of neutrino mass hierarchy and
put bounds on the mixing of $\nue$ in the ``third'' neutrino mass
eigenstate~\cite{ds,cicilia2,observ}.

When neutrinos pass through the Earth before arriving at the detector,
their spectra may get modified due to the Earth matter effects. The
presence or absence of these effects can distinguish between different
neutrino mixing scenarios~\cite{cairo}.  The comparison of neutrino
spectra at two different detectors can clearly give signatures of the
matter effects, which can be used not only for the determination of
the neutrino parameters~\cite{cicilia1,takahashi}, but also to extract
information about the density structure of the Earth
core~\cite{earthtomography}.  The measurements of the Cherenkov glow
at IceCube may also be combined with the signal at a water Cherenkov
detector like Super- or Hyper-Kamiokande to identify the Earth
effects~\cite{icecube}.

It is also possible to ascertain the presence of these matter effects
using the signal at a single detector.  It has recently been pointed
out \cite{wiggles} that the Earth matter effects on supernova
neutrinos traversing the Earth mantle give rise to a specific
frequency in the ``inverse energy'' spectrum of these neutrinos.  This
frequency, which may be identified through the Fourier transform of
the inverse energy spectrum, is independent of the initial neutrino
fluxes and spectral shapes.  Therefore, its identification serves as a
model independent signature of the Earth matter effects on SN
neutrinos, which in turn can distinguish between different scenarios
of neutrino masses and mixings.

If the SN neutrinos reach the detector ``from below,'' they have to
travel through the Earth matter. If the nadir angle of the SN
direction at the detector is less than 33$^\circ$, the neutrino path
crosses the Earth core. An investigation of the effect of the core on
the observed neutrino spectra is therefore necessary for a complete
understanding of the Earth effects. As we shall see later in this
paper, the passage through the core increases the chances of the
identification of the Earth effects. The core density is almost twice
the mantle density and this sudden density jump gives rise to new
features in the spectra.

When the SN neutrinos traverse the Earth core in addition to the
mantle, one does not get a single specific frequency as in the
mantle-only case, indeed as many as seven distinct frequencies are
present in the inverse energy spectrum.  We study the strengths of
these frequency components analytically and show that three of these
frequencies dominate.  These three frequencies can also be clearly
observed in the Fourier transform of the inverse energy spectrum when
averaged over many simulated SN neutrino signals.

Although it is difficult to isolate these frequencies individually
from the background fluctuations from a single SN burst, we suggest a
procedure that can identify the presence of these frequency components
in a sizeable fraction of cases.  Certain characteristics of the
distribution of the frequency components in the background
fluctuations are identified and used to reject the null hypothesis of
the absence of Earth effects.

We quantify the efficiency of this algorithm by simulating the SN
neutrino signal at a large scintillation detector like LENA
\cite{lena} and at a megaton water Cherenkov detector like
Hyper-Kamiokande.  Whereas the scintillation detector has the
advantage of a much better energy resolution, this is compensated in
part by the larger number of events in a megaton water Cherenkov
detector.

This paper is organized as follows. In Sec.~\ref{analytic}, we discuss
the positions and the strengths of the frequencies that characterize
the ``inverse-energy'' spectra of the neutrinos crossing the Earth
mantle as well as the core.  In Sec.~\ref{peaks}, we simulate the SN
neutrino spectra at the detectors and study the features of the peaks
with the background fluctuations averaged out.  In Sec.~\ref{bkgnd},
we introduce a method to identify the peaks in the presence of
background fluctuations and make a quantitative estimation of the
probability of peak identification as a function of the location of
the SN in the sky.  In Sec.~\ref{conclusions}, we summarize the
results.

\section{Frequencies contributed by Earth effects}
\label{analytic}

\subsection{Mixing scenarios and Earth effects}

The neutrino detectors, apart from a heavy-water detector like SNO,
can give detailed spectral information only about the $\nuebar$
flux. We shall therefore concentrate on the $\nuebar$ spectrum in this
paper.  In the presence of flavor oscillations a $\bar\nu_e$ detector
actually observes the flux
\begin{equation}
F_{\bar{e}}^D(E)  =  \bar{p}^D(E) F_{\bar{e}}^0(E) + 
\left[ 1-\bar{p}^D(E) \right] F_\xbar^0 (E)\,,
\label{feDbar}
\end{equation}
where $F_i^0$ and $F_i^D$ stand for the initial and detected flux of
$\nu_i$ respectively, and $\bar{p}^D(E)$ is the survival probability
of a $\bar\nu_e$ with energy $E$ after propagation through the SN
mantle and perhaps part of the Earth before reaching the detector. The
bulk of the $\nuebar$ are observed through the inverse beta decay
reaction $\nuebar p \to n e^+$.  The cross section $\sigma$ of this
reaction is proportional to $E^2$, making the spectrum of neutrinos
observed at the detector $N(E) \propto \sigma F_{\bar{e}}^D \propto
E^2 F_{\bar{e}}^D$.

In the absence of Earth effects, the dependence of the survival
probability on $E$ is very weak.  A significant modification of
$\bar{p}^D$ due to the Earth effects takes place only when the
neutrino mass hierarchy is normal, i.e., $m_1<m_2<m_3$, or when the
$\nue$ component of the third mass eigenstate is restricted to
$|U_{e3}|^2=\sin^2\Theta_{13} \lsim 10^{-3}$.  Here $\nu_3$ is the
neutrino mass eigenstate that has the smallest $\nue$ admixture.  The
identification of the Earth effects can then rule out the ``null
hypothesis'' of an inverted hierarchy and $|U_{e3}|^2 \gsim 10^{-3}$,
thus excluding a large chunk of the neutrino mixing parameter space.
In the language of Table~\ref{abc-table}, the Earth effects can be
present in scenarios A and C whereas they are absent in
scenario~B. Case~B is thus the null hypothesis.
 
\begin{table}[ht]
\begin{indented}\item[]
\begin{tabular}{llcc}
\hline
Case & Hierarchy &  $\sin^2 \Theta_{13}$  & Earth effects \\
\hline
A &  Normal & $\gsim 10^{-3}$  & Yes \\
B & Inverted &  $\gsim 10^{-3}$ &  No \\
C & Any & $\lsim 10^{-3}$ & Yes \\
\hline
\end{tabular}
\end{indented}
\caption{The presence of Earth effects in different neutrino mixing
  scenarios. \label{abc-table}}
\end{table}

Let us consider those scenarios where the mass hierarchy and the value
of $\Theta_{13}$ are such that the Earth effects appear for
$\bar\nu_e$. In all of these cases, $\nuebar$ produced in the SN core
travel through the interstellar space and arrive at the Earth as
$\bar{\nu}_1$. The oscillations inside the Earth are essentially
$\bar{\nu}_1$--$\bar{\nu}_2$ oscillations \cite{ds} so that we need to
solve a $2 \times 2$ mixing problem.

\subsection{Passage through only the mantle}
\label{mantle}

When the antineutrinos pass only through the mantle which has roughly
a constant density, the survival probability $\bar{p}^D$ is given by
\beq
\bar{p}^D = \left| \Big[ R(-\tm) \Phi(\phi_m) 
R(\tm-\theta_{12}) \Big]_{11} \right|^2 \,,
\label{pd-mantle}
\eeq
where $R(\theta)$ represents the $2\times 2$ rotation matrix that
rotates the neutrino state through an angle $\theta$.  Here
$\theta_{12}$ and $\tm$ are the mixing angles between $\nuebar$ and
$\bar{\nu}_2$ in vacuum and the mantle respectively.  Clearly,
$\theta_{12}$ equals the solar neutrino mixing angle.  The matrix
$\Phi(\phi_m) \equiv {\rm diag}(1,e^{i\phi_m})$ represents the change
in relative phases of $\bar{\nu}_1$ and $\bar{\nu}_2$ while traversing
the mantle: $\phi_m \approx 2 \overline{\dmsq_m} L y$, where
$\overline{\dmsq_m}$ is the mass squared difference between
$\bar{\nu}_1$ and $\bar{\nu}_2$ inside the mantle in units of
$10^{-5}$ eV$^2$, and $L_m$~is the distance traveled through the
mantle in units of 1000~km.  The ``inverse energy'' parameter is
defined as
\begin{equation}
y \equiv 12.5~\rm{MeV}/E
\label{invEparameter}
\end{equation}
where $E$ is the neutrino energy.  The energy dependence of all
quantities will always be implicit henceforth.

The survival probability Eq.~(\ref{pd-mantle}) may be written as
\begin{equation}
\bar{p}^D \approx \cos^2 \theta_{12} 
- \sin 2\tm
~ \sin (2\tm - 2\theta_{12}) 
~\sin^2 \left( \overline{\dmsq_m} L_m y \right)\,.
\label{pbar}
\end{equation}
The energy dependence of $\bar{p}^D$ introduces modulations in the
energy spectrum of $\nuebar$, which may be observed in the form of
local peaks and valleys in the spectrum of the event rate $\sigma
F_\ebar^D$ plotted as a function of $y$.  The modulations are
equispaced, indicating the presence of a single dominating frequency.
These modulations can be distinguished from random background
fluctuations that have no fixed pattern by using the Fourier transform
of the inverse energy spectrum~\cite{wiggles}.

The net $\nuebar$ flux at the detector may be 
written using Eqs.~(\ref{feDbar}) and (\ref{pbar}), in the form
\beq
F_{\bar{e}}^D =  \sin^2 \theta_{12}  F_\xbar^0 + 
\cos^2 \theta_{12} F_{\bar{e}}^0 + \Delta F^0
\bar{A}_m \sin^2(k_m y /2) \,,
\label{feDbar-y}
\eeq
where $\Delta F^0 \equiv (F_{\bar{e}}^0 - F_\xbar^0)$ depends
only on the primary neutrino spectra, whereas
$\bar{A}_m \equiv - \sin 2\tm
~ \sin (2\tm - 2\theta_{12})$ 
depends only on the mixing parameters and is independent of the
primary spectra. 
The last term in Eq.~(\ref{feDbar-y}) is the Earth oscillation term
that contains a frequency $k_m \equiv 2 \overline{\dmsq_m} L_m$ in
$y$, the coefficient $\Delta F^0 \bar{A}_m$ being a relatively slowly
varying function of $y$.  The first two terms in Eq.~(\ref{feDbar-y})
are also slowly varying functions of $y$, and hence contain
frequencies in $y$ that are much smaller than $k_m$.  The dominating
frequency $k_m$ is the one that appears in the modulation of the
inverse-energy spectrum.

The frequency $k_m$ is completely independent of the primary
neutrino spectra, and indeed can be determined to a good accuracy from
the knowledge of the solar oscillation parameters, the Earth matter 
density, and the position of the SN in the sky.

\subsection{Passage through the mantle and the core}
\label{also-core}

We now study analytically the spectral modulations arising when the
neutrinos travel through both the mantle and the core, and the effect
of the sharp density jump at their boundary.  We denote the mixing
angles, phases and mass squared differences in the core by replacing
the superscript/subscript ``$m$'' in the last section by ``$c$''.  The
neutrinos cross two sections of the mantle with equal length $L_m/2$
each.  We denote the total distance traveled through the core by
$L_c$.

The anti-neutrino survival probability is given by
\barr
\bar{p}^D & = & \left| \Big[ R(-\tm) \Phi(\phi_m/2) 
R(\tm - \tc) \Phi(\phi_c) \right. \nonumber \\
& & \phantom{\Big[ R(-\tm)\Phi(\phi_m/2)} 
\left. \times R(\tc - \tm) \Phi(\phi_m/2) 
R(\tm-\theta_{12}) \Big]_{11} \right|^2 \,.
\label{pd-core}
\earr
This may be written in the form
\beq
\bar{p}^D \approx \bar{A}_0 + \sum_{i=1}^7 \bar{A}_i
\sin^2(\phi_i/2) \,,
\label{pdsum}
\eeq 
where $\bar{A}_0\equiv\cos^2 \theta_{12}$.  The explicit expressions
for the other $\bar{A}_i$ and $\phi_i$ are given in
Table~\ref{tab:7freq}.  As in the mantle-only case, $\phi_m \equiv
2\dmbarm L_m y$ and $\phi_c \equiv 2\dmbarc L_c y$.  Note that in the
absence of travel through the core, $\phi_c=0$ and Eq.~(\ref{pdsum})
reduces to Eq.~(\ref{pbar}).

\begin{table}[ht]
\begin{indented}\item[]
\begin{tabular}{|c|c|c|c|}
\hline
& & & \\
$i$ & $\phi_i$ &  $\bar{A}_i$ &  \\
\hline
& & & \\
1 & 
$\phi_m/2$ & $-\frac{1}{2} \sin(2\theta_{12}-4\tm) \sin(4\tc - 4\tm)$ &
${\cal O}(\omega)$ \\
& & & \\
2 &
$ (\phi_m/2+\phi_c)$ &
$\cos^2(\tc-\tm)\sin(2\theta_{12}-4\tm)\sin(2\tc-2\tm)$ & ${\cal O}(\omega)$\\ 
& & & \\
3 &
$ (\phi_m+\phi_c)$ & $ \sin(2\theta_{12}-2\tm) \cos^4(\tc-\tm)
\sin (2\tm)$ &
${\cal O}(\omega)$ \\
& & & \\
4 &
$ \phi_c$ & $- \sin^2(2\tc-2\tm)
~[\cos(2\theta_{12}-4\tm)- $ & \\
& &
$-\frac{1}{2}\sin(2\theta_{12}-2\tm)\sin(2\tm)]$ &
${\cal O}(\omega^2)$ \\
& & & \\
5 &
$ \phi_m$ & $\frac{1}{2} \sin(2\theta_{12}-2\tm) \sin^2(2\tc - 2\tm)
\sin (2\tm)$ &
${\cal O}(\omega^3)$ \\
& & & \\
6 &
$ (\phi_m/2-\phi_c)$ & $-2 \sin(2\theta_{12}-4\tm) \cos(\tc-\tm)
\sin^3(\tc-\tm)$ &
${\cal O}(\omega^3)$ \\
& & & \\
7 &
$ (\phi_m-\phi_c)$ & $ \sin(2\theta_{12}-2\tm) \sin^4(\tc-\tm)
\sin (2\tm)$ &
${\cal O}(\omega^5)$ \\
\hline
\end{tabular}
\end{indented}
\caption{Explicit expressions for $\phi_i$ and $\bar{A}_i$ in
Eq.~(\ref{pdsum})
\label{tab:7freq}}
\end{table}

The net $\nuebar$ flux at the detector may be 
written using Eqs.~(\ref{feDbar}) and (\ref{pdsum}) in the form
\beq
F_{\bar{e}}^D =  \sin^2 \theta_{12}  F_\xbar^0 + 
\cos^2 \theta_{12} F_{\bar{e}}^0
+ \Delta F^0
\sum_{i=1}^7 \bar{A}_i \sin^2(k_i y /2) \,,
\label{fesum-y}
\eeq
where $k_i \equiv \phi_i/y$ are the dominating frequencies.

\begin{figure}[ht]
\begin{center}
\hbox{\epsfig{file=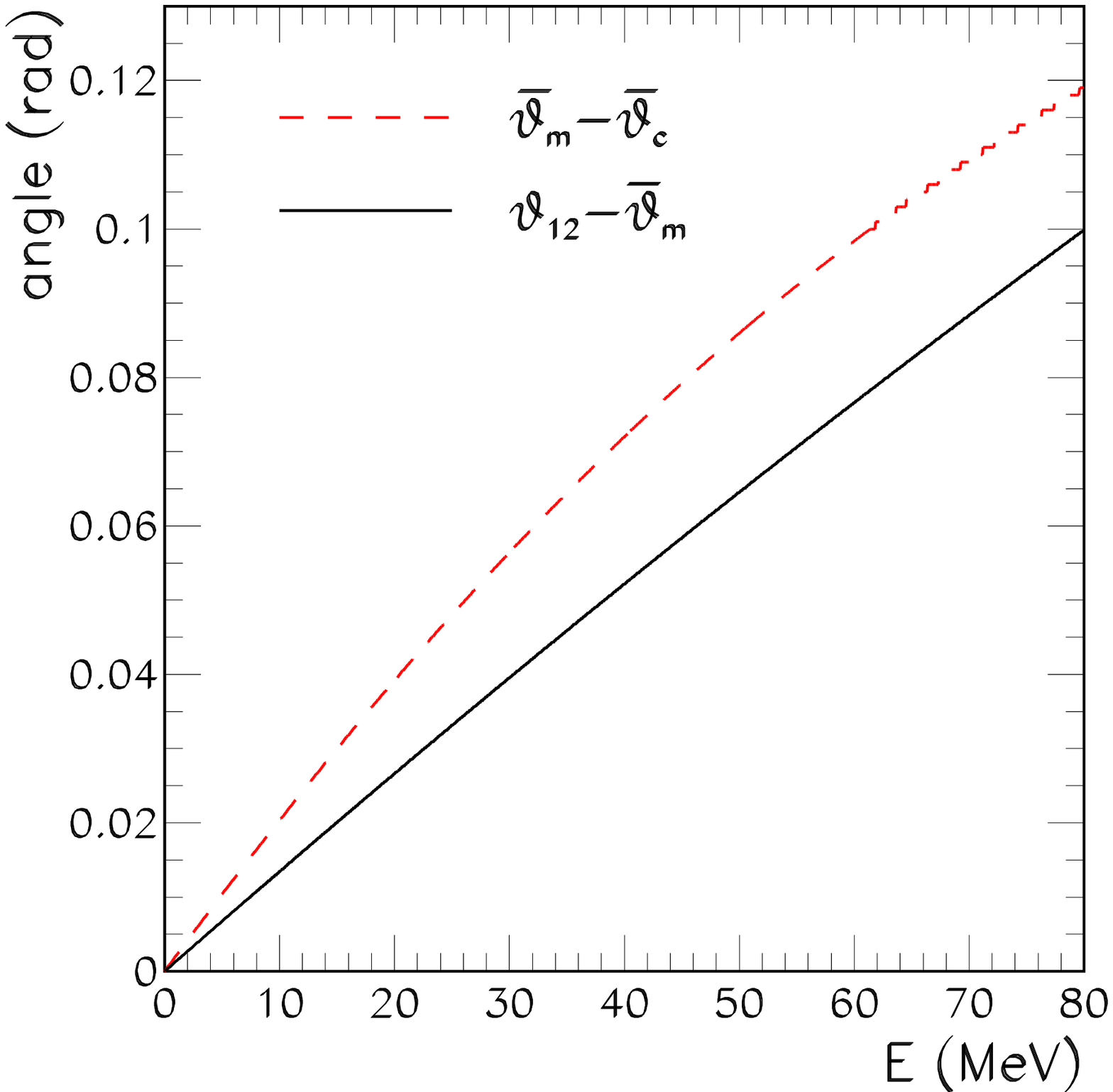,width=8.cm,angle=0}
\epsfig{file=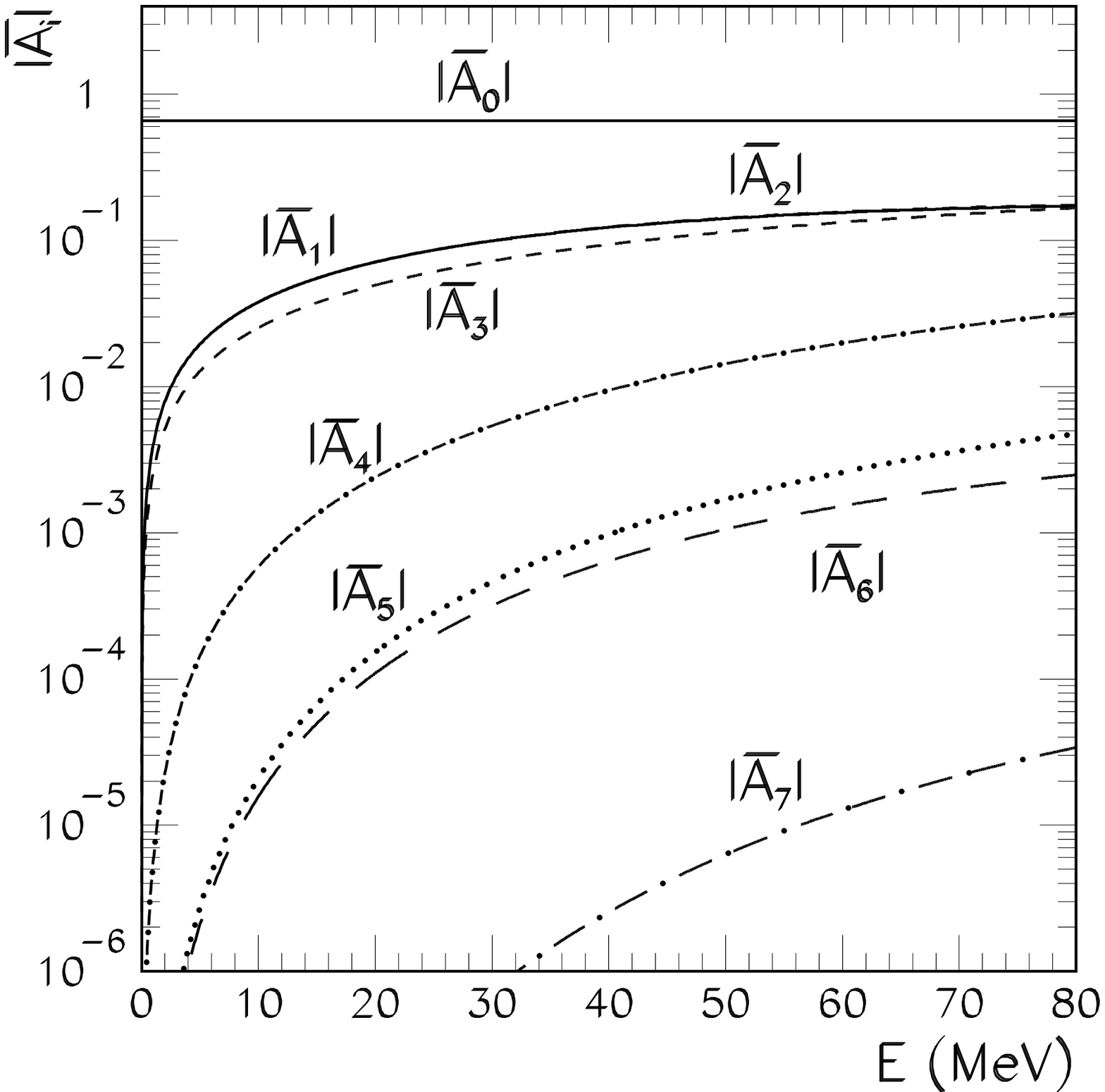,width=8.cm,angle=0}}
\end{center}
\caption{ Left: The energy dependence of the changes in the neutrino
mixing angles during the passage through vacuum, mantle and
core. Right: The energy dependence of the magnitude of $\bar{A_i}$'s in
Eq.~(\ref{fesum-y}).
\label{fig:Ais}}
\end{figure}

Not all the frequencies $k_i$ are equally important.  We estimate the
relative magnitudes of these terms in the following manner.  The
mixing angles in the mantle and the core, $\tm$ and $\tc$
respectively, are given by
\beq
\sin^2 2\bar{\theta}_{m(c)} = \frac{\sin^2 2\theta_{12}}
{\sin^2 2\theta_{12} + (\cos 2\theta_{12}- 2EV_{m(c)}/\Delta m^2)^2 }\,,
\eeq
where $V_{m(c)}$ is the effective potential due to the matter in the
mantle (core) for $\nuebar$.  For the densities of the mantle as well
as the core, both $(\theta_{12}-\tm)$ and $(\tm-\tc)$ are small
numbers of order 0.1 as can be seen in the left panel of
Fig.~\ref{fig:Ais}.  In the last column of Table~\ref{tab:7freq} we
symbolically denote either of these quantities by $\omega\approx 0.1$
and show the power of $\omega$ involved in the coefficient of that
particular frequency term.  Terms with higher powers of $\omega$ are
suppressed.

We observe that the low-frequency component $\bar{A}_0$, that does not
contribute to the Earth effects, is the largest in magnitude.  Among
the terms relevant for the Earth effects there are three dominant
frequencies, corresponding to the first three terms in the summation
in Eq.~(\ref{fesum-y}).  The fourth term is subleading and the rest
are too suppressed to be of any significance. The right panel of
Fig.~\ref{fig:Ais} confirms that the first three terms have similar
magnitudes, in particular $|\bar{A}_1|\approx |\bar{A}_2|$, whereas
the others are significantly suppressed for all the energies relevant
for SN neutrinos.  We expect the first three terms to give rise to
three dominant peaks in the Fourier spectrum of the inverse energy
spectrum.  Note that since $\dmbarc \approx \dmbarm \approx
\dmsq_\odot$ to within 20\% in the relevant parameter range, the
positions of these peaks are also known once the distance traversed
through the Earth is known, independently of the primary neutrino
spectra.  This distance can be determined with sufficient precision
even if the SN is optically obscured using the pointing capability of
neutrino detectors~\cite{pointing}.

\newpage

\section{Peaks in the power spectrum of  $\mathbf{\nuebar}$}
\label{peaks}

\subsection{Definitions}
\label{definitions}

The neutrino signal is observed as a discrete set of events.
The measured energy of these events is correlated
to the neutrino energy via
kinematics and detector properties. Then 
we define the power spectrum of $N$ detected events as
\beq
G(k) \equiv \frac{1}{N} \left| \sum_{i=1}^N e^{iky_i} \right|^2 \,.
\label{ft-def}
\eeq
In the absence of Earth effect modulations, $G(k)$ is expected to
have an average value of one for $k \gsim 40$ \cite{wiggles}. 
Earth effects introduce
peaks in this power spectrum at specific frequencies, the identification
of which correspond to the identification of the Earth effects.

Before defining an algorithm to analyze neutrino signals from a
single SN, we perform a check of the analytical features derived in
Sec.~\ref{analytic} with a realistic Monte Carlo simulation.
For the time-integrated neutrino fluxes we assume distributions
of the form~\cite{Keil:2002in}
\begin{equation}\label{eq:spectralform}
F^0=
\frac{\Phi_0}{E_0}\,\frac{(1+\alpha)^{1+\alpha}}{\Gamma(1+\alpha)}  
\left(\frac{E}{E_0}\right)^\alpha 
\exp\left[-(\alpha+1)\frac{E}{E_0}\right] \,,
\end{equation}
where $F^0$ denotes the flux of a neutrino species emitted by the SN
scaled appropriately to the distance traveled from the SN to Earth.
Here $E_0$ is the average energy and $\alpha$ a parameter 
that relates to the width of the spectrum and
typically takes on values 2.5--5, depending on the flavor and the
phase of neutrino emission. The values of the total flux $\Phi_0$ and
the spectral parameters $\alpha$ and $E_0$ are generally different for
$\nu_e$, $\bar\nu_e$ and $\nu_x$, where $\nu_x$ stands for any of
$\nu_{\mu,\tau}$ or $\bar\nu_{\mu,\tau}$.  

\begin{table}[ht]
\begin{indented}\item[]
\begin{tabular}{cccccc}
\hline
Model & $\langle E_0(\nu_e) \rangle$ & $\langle E_0(\nuebar) \rangle$&
$\langle E_0(\nux) \rangle$ & {\large $\frac{\Phi_0(\nu_e)}{\Phi_0(\nu_x)}$} &
{\large $\frac{\Phi_0(\nuebar)}{\Phi_0(\nu_x)}$}\\
\hline
G & 12 & 15 & 18 & 0.8 & 0.8 \\
L & 12 & 15 & 24 & 2.0 & 1.6 \\
\hline
\end{tabular}
\end{indented}
\caption{The parameters in the neutrino spectra models motivated from
  SN simulations of the
Garching (G) group and the Livermore (L) group. We assume $\alpha=3$ for
all neutrino species.
\label{tab:models}}
\end{table} 

In order to study the model dependence, we consider two models that
give very different predictions for the neutrino spectra.  The first
is motivated by the recent Garching calculation \cite{garching} that
includes all relevant neutrino interaction rates, including nucleon
bremsstrahlung, neutrino pair processes, weak magnetism, nucleon
recoils and nuclear correlation effects.  The second is the result
from the Livermore simulation \cite{livermore} that represents
traditional predictions for flavor-dependent SN neutrino spectra that
have been used in many previous analyses.  The parameters of these
models are shown in Table~\ref{tab:models}.  To study the background,
we use the mixing parameters of scenario B in Table~\ref{abc-table} in
which the Earth effects are absent.

For the Earth density profile we use a 5-layer approximation of the
PREM model as parametrized in Ref.~\cite{lisi-profile}.  We start with
a SN at 10 kpc and simulate the neutrino signal at a 32 kiloton
scintillation detector and a megaton water Cherenkov detector.  In
both detectors, the neutrino signal is dominated by the inverse beta
reaction $\nuebar p \to n e^+$, while all other reactions have a
negligible influence on the analysis below.  The detector response is
taken care of in the manner described in Refs.~\cite{observ,pointing}.
The major difference between the scintillation and the water Cherenkov
detector is that the energy resolution of the scintillation detector
is roughly six times better than that of the Cherenkov detector.  This
compensates the size advantage of a megaton water Cherenkov detector.

\subsection{Large scintillation detector}
\label{scin}

\begin{figure}[b]
\begin{indented}\item[]
\epsfig{file=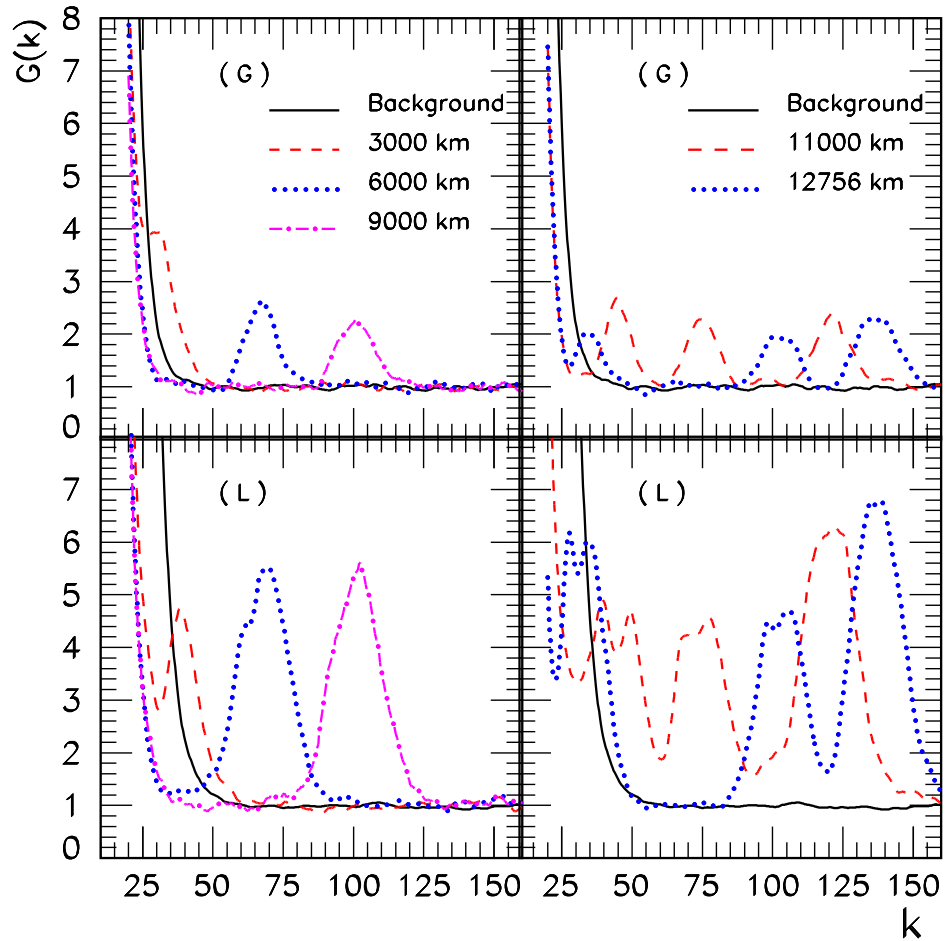,width=5.5in}
\caption{Averaged power spectra in the case of a large scintillator
  detector for different SN models, Garching (G) and Livermore (L),
 and distances travelled through the
  Earth.
\label{fig:Lena_G-L_av_mantle-core_bck}}
\end{indented}
\end{figure}

To start with, we consider the power spectrum resulting from averaging
1000 SN simulations. This eliminates the fluctuations in the
background, and illustrates the characteristics of the peaks in a
clear manner.  The power spectrum at a 32~kt scintillation detector
for different distances traveled through the Earth is shown in
Fig.~\ref{fig:Lena_G-L_av_mantle-core_bck}.  The top panels use the
Garching model whereas the bottom ones use the Livermore model.  The
left panels show three typical cases when the neutrinos traverse only
the mantle, whereas the right panels represent the passage of
neutrinos through the mantle as well as the core.  Only inverse beta
decay events have been taken into account. We have checked that the inclusion of the
other reactions like the elastic scattering off electrons or the
charged current reactions on carbon do not change the results. The
neutral current reactions on carbon, although providing a large number
of events, have been neglected, as the monoenergetic photon produced by
the decay of the excited carbon could be tagged in such a
detector~\cite{Cadonati:2000kq}.

\bigskip

\noindent The following observations may be made: 
\begin{enumerate}
\item
The average background approaches 1 for $k \gsim 40$, as expected
\cite{wiggles}. The region $k \lsim 40$ is dominated by the
``0-peak,'' which is a manifestation of the low frequency terms in
Eqs.~(\ref{feDbar-y}) and (\ref{fesum-y}).  Note that the 0-peak in
the background case is wider than that in the signal case. This is
because the background case corresponds to the scenario B, which is
also the one wherein there is a complete interchange of the $\nuebar$
and $\nuxbar$ spectra.  The energy of the detected $\nuebar$ spectrum
is thus higher, which results in a broader 0-peak.

\item
When the neutrinos traverse only the mantle, only one peak appears at
the expected value of $k_m$ that is proportional to the distance $L_m$
traveled through the mantle.  For $L_m < 3000$ km, the position of the
peak lies at such low frequencies that it can hardly be distinguished
from the 0-peak.  This illustrates that neutrinos must travel a
minimum distance through the Earth before the Earth effects become
observable.

\item
When the neutrinos travel also through the core, we observe three
dominant peaks in each case, corresponding to $k_1,k_2$ and $k_3$ in
Eq.~(\ref{fesum-y}).  We observe that, as the total distance traveled
through the Earth increases,
\begin{itemize}
\item the third peak moves to higher $k$, since $k_3 \propto 
(L_m + L_c)$.
\item the second peak, whose position is proportional to
   $(L_m/2+L_c)$, also gets shifted towards higher $k$ as the
   increase in $L_c$ is larger than the decrease in $L_m/2$.
\item the first peak, on the other hand, moves to lower $k$ values, 
since as the trajectory of neutrinos approaches the center of 
the Earth, the distance traveled through the mantle decreases
and so does the frequency of the lowest peak, $k_1 \propto L_m/2$.
This makes the detection of this peak harder at larger distances 
traveled through the Earth.
\end{itemize}

\item
The model independence of the peak positions may be confirmed by
comparing the top and bottom panels of
Fig.~\ref{fig:Lena_G-L_av_mantle-core_bck}.  The peaks obtained with
the Livermore model are stronger as a result of the larger difference
between the $\nuebar$ and $\nuxbar$ spectra in that model, which
increases the value of $\Delta F^0$ in Eqs.~(\ref{feDbar-y}) and
(\ref{fesum-y}).  However, the positions of the peaks are the same as
those obtained with the Garching model.
\end{enumerate}

\subsection{Megaton water Cherenkov detector}
\label{cherenkov}

\begin{figure}[b]
\begin{indented}\item[]
\epsfig{file=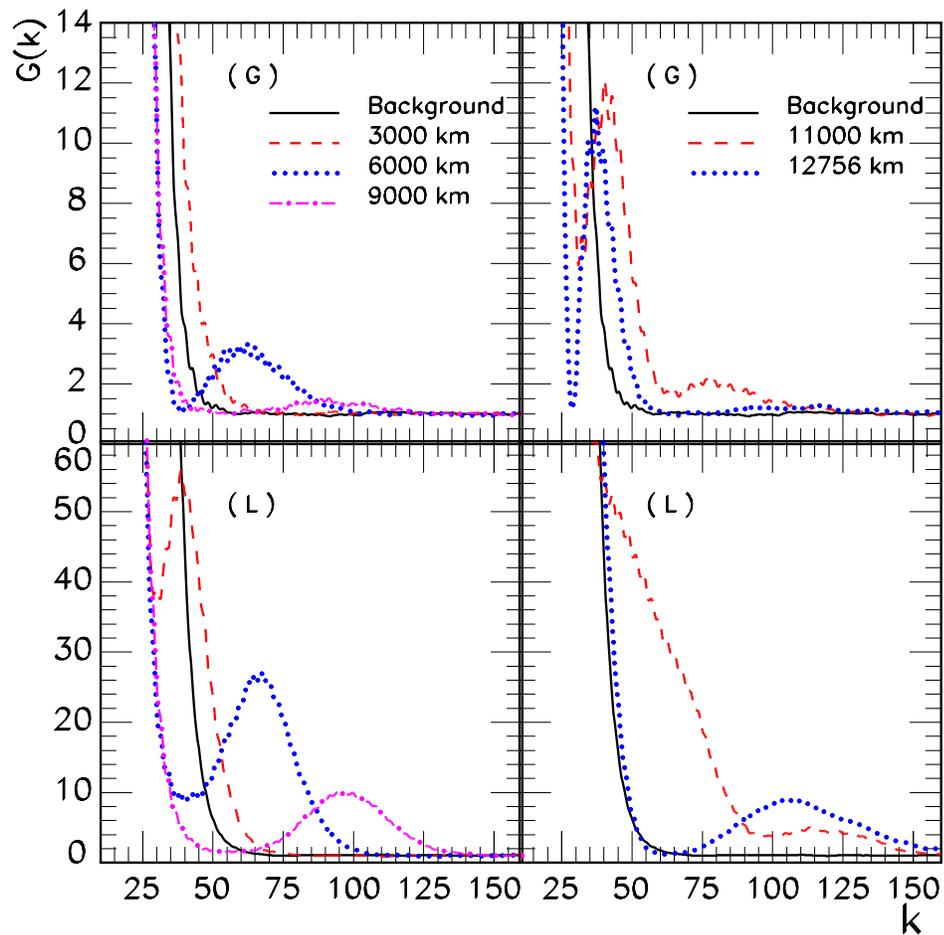,width=5.5in}
\end{indented}
\caption{Same as the Fig.~\ref{fig:Lena_G-L_av_mantle-core_bck} but
  for the case of HK. 
\label{fig:HK_G-L_av_mantle-core_bck}}
\end{figure}

The energy resolution of a water Cherenkov detector is about a factor
of six worse than that of a scintillation detector. This means that
the energy spectrum is more ``smeared out'' and higher frequencies in
the spectrum are more suppressed.  This makes the peak identification
more difficult, and even a detector of the size of Super-Kamiokande
turns out not to be sufficient \cite{wiggles}.  We show the power
spectrum expected at a megaton water Cherenkov detector in
Fig.~\ref{fig:HK_G-L_av_mantle-core_bck} for the two SN models
considered here and for different locations of the SN.  Again we have
only assumed the events from the inverse beta decay, since the
contribution from other reactions is significantly smaller.

\bigskip

\noindent The following observations may be made from the power spectra:
\begin{enumerate}
\item
The 0-peak is broader, which makes 
the averaged background approach one for larger values of $k$
($k\simeq 50$) than
in the case of the scintillation detector. 
 This is because the energy smearing decreases the strengths
of high frequency components and increases the strengths of low
frequency components.
\item
In the case of neutrino propagation only through the mantle, the peak
shifts to higher $k$ as $L_m$ increases, as expected.  However, the
suppression of high frequencies tends to shift the peak locations to
slightly lower $k$ values than at the scintillation detector.  In
addition, as the peak position moves to larger $k_m$, the strength of
the peak decreases and the peak becomes harder to detect.
\item
When the neutrinos traverse both the mantle and the core,
it is observed that
\begin{itemize}
\item the third peak among the expected three dominant ones,
corresponding to $k_3 \propto (L_m + L_c)$, is highly 
suppressed due to large $k_3$ and is undetectable.
\item the other two peaks corresponding to $k_1 \propto L_m/2$
and $k_2 \propto (L_m/2 + L_c)$ have lower $k$ values, and
are not as suppressed as the $k_3$ peak. Moreover, the $k_1$
peak is stronger than the $k_2$ one.
\item the $k_1$ peak moves to lower frequencies as the total
distance traveled through the Earth increases. 
Beyond a certain distance, it merges with the background
0-peak and becomes undetectable.
\end{itemize}
\item
The peak positions with the Garching and Livermore spectra
are at nearly the same positions, though the extra feature of
high $k$ suppression is observed to shift the peaks with the
Garching model to slightly lower values of $k$ as compared to those 
with the Livermore model. The peaks with the Livermore model 
naturally have more strength than the ones with the Garching model.
\end{enumerate} 

\section{Distinguishing the peaks from the background}
\label{bkgnd}

\subsection{An algorithm for peak identification}
\label{algorithm}

Though the analytic approximations seem to work well with the averaged
power spectrum, the understanding of the statistical fluctuations
within the signal from a single SN is crucial for the identification
of the peaks.  As observed in \cite{wiggles} and confirmed in
Sec.~\ref{peaks}, the average of the background power spectrum is
indeed one for all values of $k$ after the dominant low frequency peak
in the power spectrum dies out. As long as we are free of the
influence of this low frequency peak, the area under the power
spectrum between two fixed frequencies $k_{\rm min}$ and $k_{\rm max}$
is on an average $(k_{\rm max}-k_{\rm min})$. In the absence of Earth
effects, this area will have a distribution centered around this mean.
The Earth effect peaks tend to increase this area. If the area in a
specific interval is found to be more than what mere background
fluctuations can allow, the peak can be identified with confidence.

\begin{figure}[b]
\begin{center}
\hbox{\epsfig{file=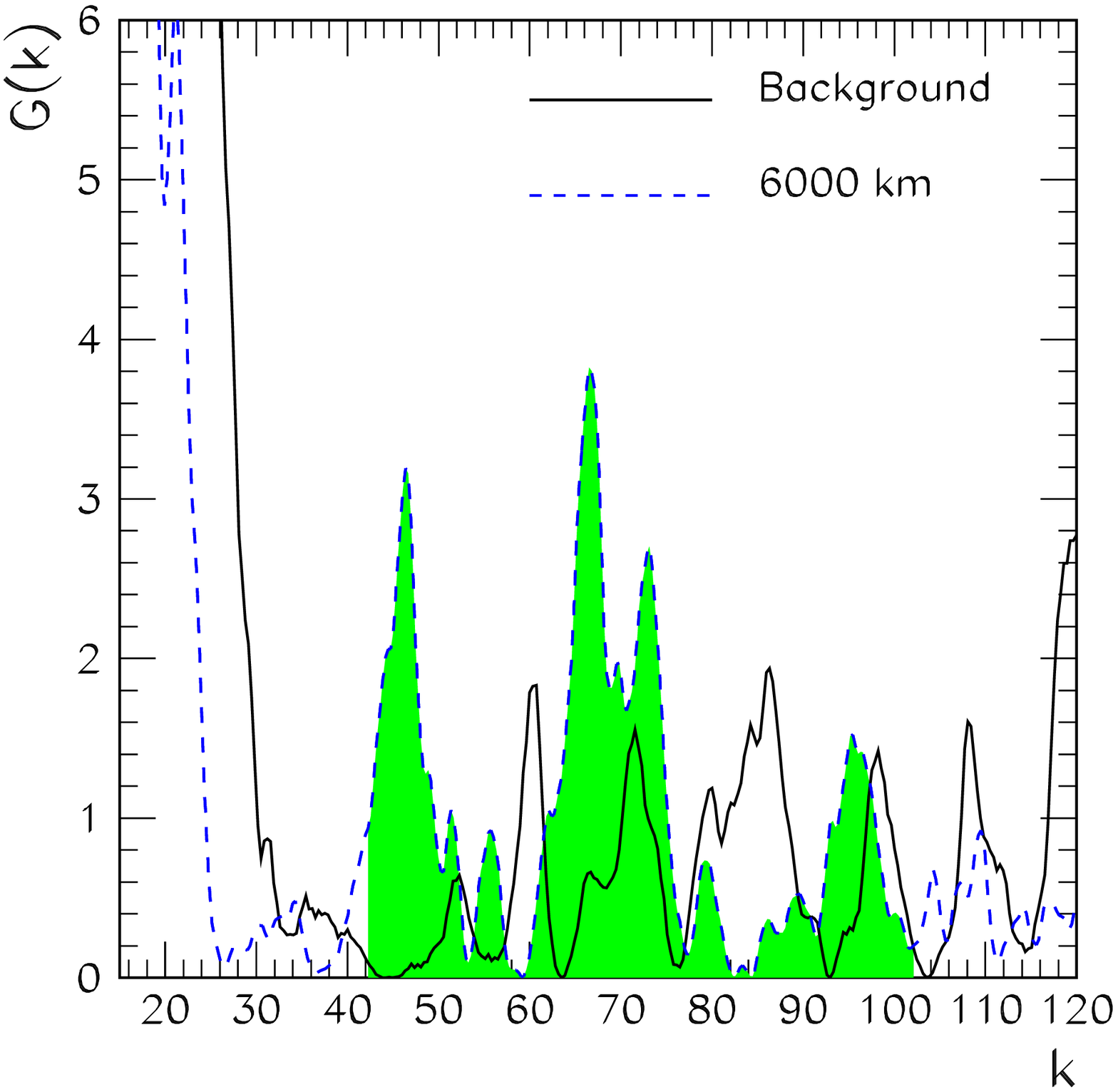,width=8cm}
\epsfig{file=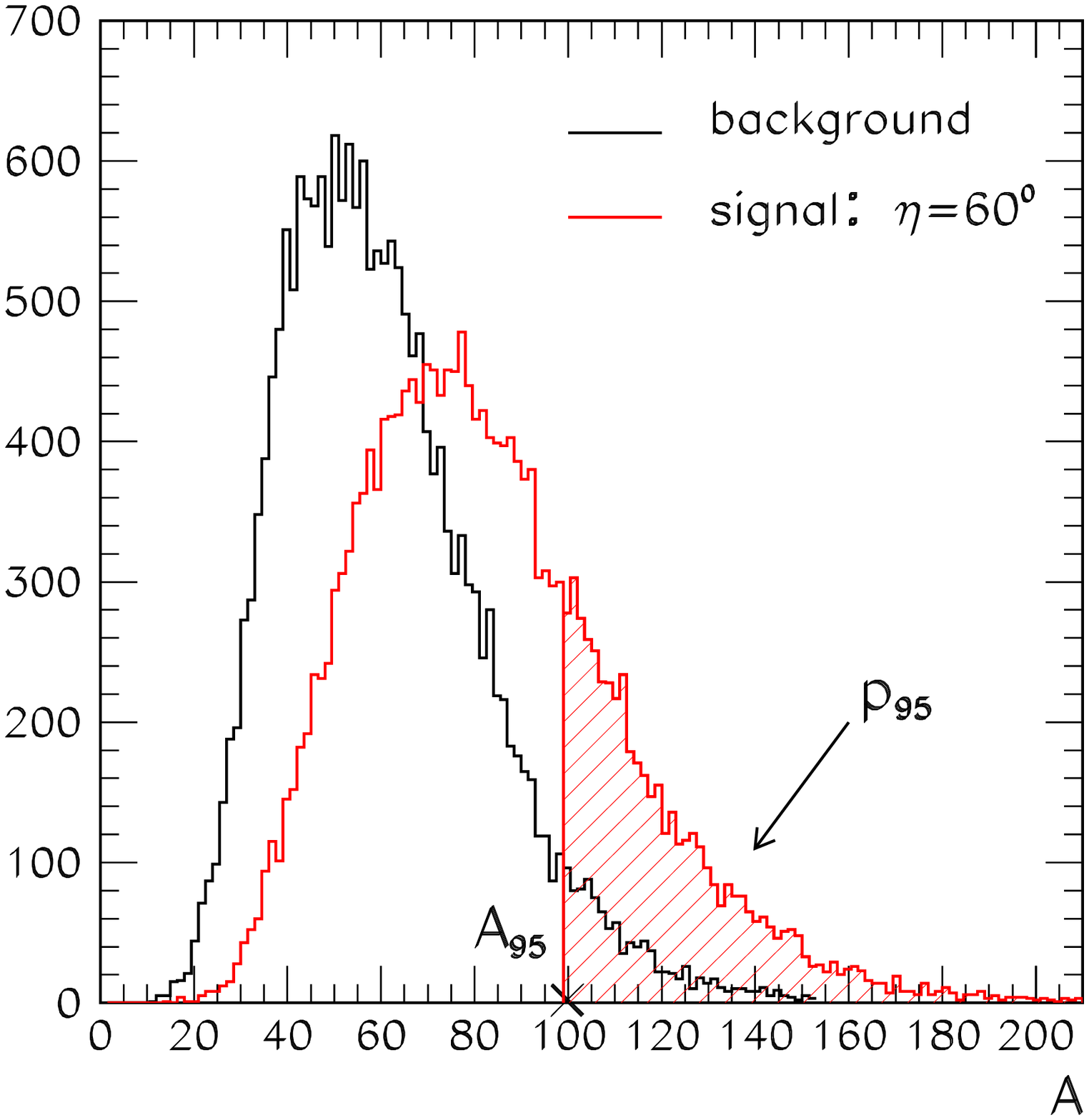,width=8cm}}
\end{center}
\caption{Left: Realistic spectrum from a single simulation.
Right: Area distribution of the background (black) and the signal (red)
  obtained for a 32 kton scintillator detector and Garching model for $\eta=60$.
\label{fig:exemple1-3}}
\end{figure}

Since in the real world the presence of fluctuations in the signal
will spoil any naive theoretical peak, we need to
introduce a prescription to carry out the analysis.
\begin{enumerate}
\item[1.]  Once we know the total distance traveled by the neutrinos
through the Earth, we can calculate the position where the peak should
lie. Instead of looking for the maximum in the height of the power
spectrum, we consider a more robust observable, namely the area around
the position of the peak, as illustrated in Fig.~\ref{fig:exemple1-3}.
When only one peak centered at $k_m$ is expected, we consider the
interval $k_m\pm \Delta k$ with $\Delta k=30$, roughly the expected
width of the peak.  In order to avoid the 0-peak contamination we set
a lower limit at $k=40$ ($k=50$) for the scintillator (water
Cherenkov) detector.
When the neutrinos also cross the Earth core, multiple peaks are
present and we measure the area from $k=40$ until $k=160$ in such
cases.

\item[2.]
The next step is to analyze the statistical significance of the
result obtained. For this purpose we compare the value of the
measured area with the distribution of the area in  the case of no
Earth matter effects. 
Since the different frequencies are not uncorrelated, the background
distribution is not simply Gaussian centered at 
$A_{\rm av} = (k_{\rm max}-k_{\rm min})$ and with width
$\sqrt{k_{\rm max}-k_{\rm min}}$. Therefore, we perform a Monte Carlo
analysis of the background case and calculate the exact distribution 
with which one can compare the actual area measured.
We illustrate this in the right panel of 
Fig.~\ref{fig:exemple1-3}, where the black curve
shows the area distribution of the background.
The confidence level of peak identification may then be
defined as the fraction of the area of the background distribution
that is less than the actual area measured. 
We denote the area corresponding to $\alpha$\% C.L. by $A_\alpha$.
Figure~\ref{fig:exemple1-3} also shows $A_{95}\approx 100$, the area
corresponding to a peak identification with 95\% confidence.
\end{enumerate}

\subsection{Quantifying the efficiency of the algorithm}
\label{efficiency}

Since the peak identification algorithm is statistical in nature,
it is worthwhile to have an idea of the probability with which
a peak can be identified with a given confidence. 
This probability clearly depends on the distance traveled by the 
neutrinos through the Earth, which in turn is determined by the
location of the SN in the sky.
We parameterize the SN location by the nadir angle $\eta$ of the
SN direction at the detector. 

\begin{figure}[b]
\begin{center}
\hbox{\epsfig{file=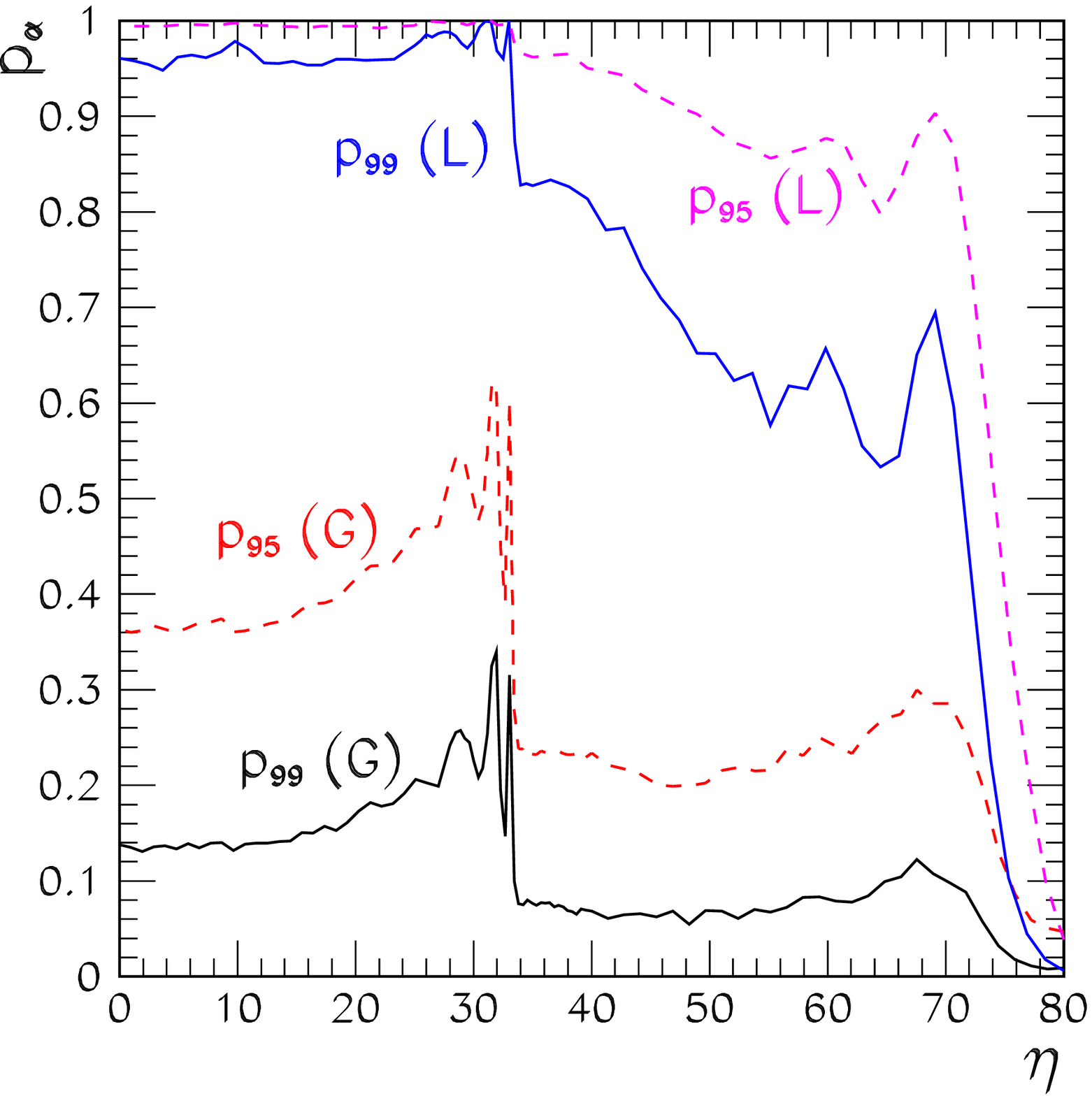,width=8cm}
\epsfig{file=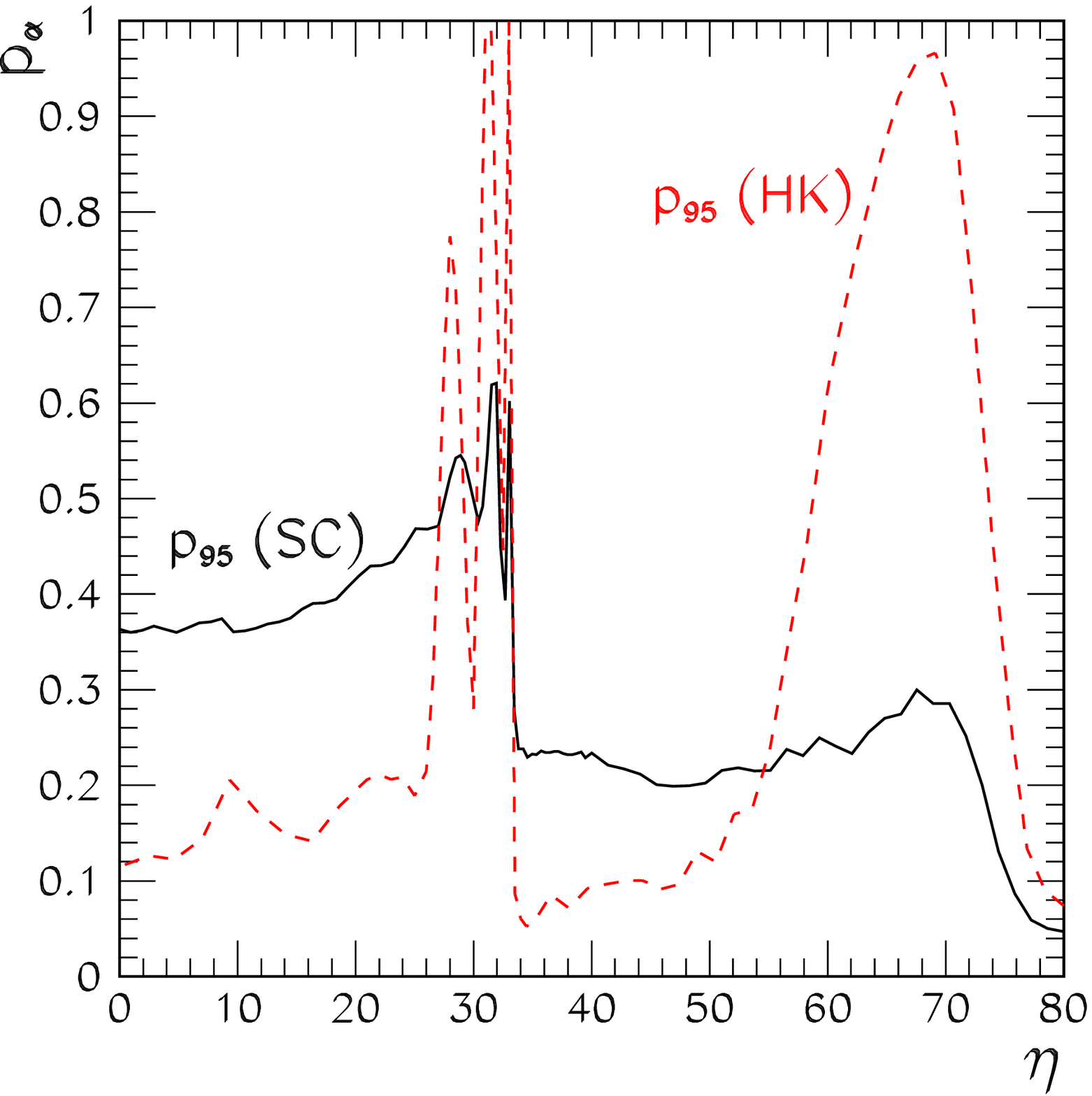,width=8cm}}
\end{center}
\caption{Left: Comparison of $p_{95}$ and $p_{99}$ for the Garching
  (G) and Livermore (L) SN models in a 32 kton scintillator detector. 
Right: Comparison 
   of $p_{95}$ in this large scintillator detector (SC) and 
in the case of a megaton water Cherenkov (HK), for the Garching model.
\label{fig:Lena_G-L_eta}}
\end{figure}

We simulate the area distribution for the ``signal'' using the
neutrino mixing scenarios that allow Earth effects and compare it with
the background distribution.  The probability $p_\alpha$ of peak
identification at $\alpha$\% C.L.\ is the fraction of the area of the
signal distribution above $A_\alpha$. In Fig.~\ref{fig:exemple1-3}, it
has been indicated with the red-hashed region.
In Fig.~\ref{fig:Lena_G-L_eta} we show $p_{95}$ and $p_{99}$
as a function of $\eta$ in the case of a  scintillation detector, for
the two SN models considered. An increase in the distance 
traveled through the Earth corresponds to a decrease in $\eta$.
The passage through the core corresponds to $\eta < 33^\circ$.
One can see that the presence of the core enhances the chances of
detecting the Earth matter effects.  The probability is higher at
values of $\eta$ close to the boundary between the mantle and the core
because the three peaks are clearly visible.  The oscillation pattern
arising at this region stems from the interference of the first two
peaks in the spectrum, whose positions at this point differ only by
$\phi_c\propto L_c$.  As the distance traveled by the neutrinos
through the core increases, $p_\alpha$ decreases, the reason being the
approach of the first peak to the lower limit frequency, $k\simeq
k_{\rm min}$, and its eventual disappearance.

As expected, the chances of peak identification are also higher when
the primary spectra of $\nuebar$ and $\nuxbar$ differ more. As shown
in Fig.~\ref{fig:Lena_G-L_eta}, the Livermore model predicts much
larger chances of a successful peak identification.

In the right panel of Fig.~\ref{fig:Lena_G-L_eta}, we assume the
Garching model and compare the results obtained with a 32 kton
scintillator detector and a megaton water Cherenkov detector.  In the
latter case, as neutrinos travel more and more distance in the mantle
the peak moves to higher $k$ values, and due to the high $k$
suppression as described in Sec.~\ref{cherenkov}, the efficiency of
peak identification decreases. However, when the neutrinos start
traversing the core, additional low $k$ peaks are generated and the
efficiency increases again.

\begin{figure}
\begin{indented}\item[]
\epsfig{file=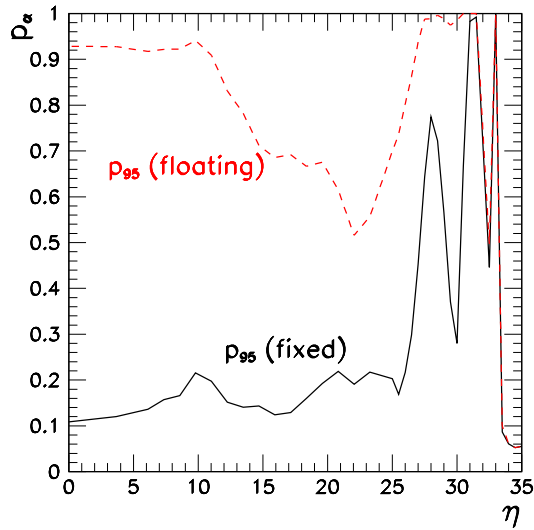,width=3.0in}
\end{indented}
\caption{Comparsion of $p_{95}$ for two different methods of
  integration, with {\em fixed} and {\em floating} lower limit, in the
  case of a megaton water Cherenkov detector and the Garching model.
\label{fig:HK_G_eta_method}}
\end{figure}

One of the features of this algorithm is its robustness. However
in some cases it turns out to be very conservative. For instance,
if we have a look at the megaton water Cherenkov,
the peaks at the lowest frequencies are almost eaten-up by the 0-peak
of the background when the neutrinos cross the Earth core,
cf.~Fig.~\ref{fig:HK_G-L_av_mantle-core_bck}.  Setting $k_{\min}=50$
as the lower limit of the area integration results therefore in a loss of
considerable amount of information under these conditions. 
In this particular case it is possible to optimize the efficiency of
the method by choosing a {\em floating\/} lower cut: instead of
considering a fixed value, $k_{\rm min}=50$, as the lower limit for the area
integration, one defines $k_{\rm min}$ as the frequency 
at which the spectrum has the first minimum after neglecting the
effect of spurious fluctuations.
With this modified prescription one can again compare the area
distribution for the background and that of the signal, and calculate
a new $p_\alpha$. 
We have checked this method for both the scintillator and the water
Cherenkov detector. 
For the former the improvement is not relevant. The reason is that 
the second and third peaks contribute significantly to the signal. So, 
even when the first peak
disappears behind the 0-peak of the background for fixed lower
cut-off, the loss of information is not important.
However in the case of a water Cherenkov detector the efficiency is
significantly enhanced for  paths traversing the Earth core 
as can be seen in Fig.~\ref{fig:HK_G_eta_method}. 
Due to the suppression of the peaks located at higher frequencies only
the first peak contributes significantly to the signal. However for
trajectories involving small $\eta$ this peak is centered at very low
frequencies, almost completely hidden by the 0-peak of the
background. Under this situation if one allows the lower bound to
shift to smaller frequencies the whole peak contributes to the signal,
and therefore the probability to see the Earth matter effects increases.
On the other hand when the neutrinos only cross the mantle the
location of the unique peak is mostly at $k>k_{\rm
  min}=50$. Therefore, this modified prescription does not help much
to improve the efficiency to observe the modulation of the neutrino
spectra due to the Earth matter effects.

\section{Summary and conclusions}
\label{conclusions}

When neutrinos coming from a core-collapse supernova pass through the
Earth before arriving at the detector, the spectra may get modified
due to the Earth matter effects.  The presence or absence of these
effects can distinguish between different neutrino mixing scenarios.
We have seen that these Earth matter effects on supernova neutrinos
can be identified at a single detector through peaks in the Fourier
transform of their ``inverse energy'' spectrum.
The position of these peaks are independent of
the initial neutrino fluxes and spectral shapes.

We have performed an analytical study of the positions and the
strengths of the frequencies that characterize the inverse-energy
spectrum of the neutrinos for different neutrino trajectories through
the Earth.
In the case that the SN neutrinos only traverse the mantle a single
peak shows up in the power spectrum. In contrast we have observed that
if both the mantle and the core are crossed before the neutrinos reach
the detector as many as seven distinct frequencies are present in the
inverse energy spectrum. However only three peaks are dominant 
in the  power spectrum. This increase in the number of
expected peaks leads to an easier identification of the Earth matter
effects.

In order to illustrate the qualitative features of the present
analysis we have considered the power spectrum resulting from
averaging 1000 SN simulations for different SN models and different
detector capabilities.
In particular we have assumed a 32 kton scintillator detector and a
megaton water Cherenkov detector.  We have shown how the energy
resolution turns out to be crucial in detecting the modulation
introduced in the neutrino spectra by the Earth matter effects.
First, the better resolution of the scintillator detector compensates
for the larger water Cherenkov detector size.  On the other hand, the
worse energy resolution in water Cherenkov detectors does not only
imply the need of a larger volume but also suppresses significantly
the peaks at higher frequencies, in contrast to the case of
scintillator detectors.

We have considered two
different SN models as an illustration of the current uncertainties in the
initial fluxes. We have observed that the strength of the peaks is
larger in those SN models with bigger differences between $\bar\nu_e$
and $\bar\nu_\mu$ spectra. However, we have found that the position of
the peaks is model independent.  Therefore their identification serves
as a clear signature of the Earth matter effects on SN neutrinos,
which in turn can help to discard the neutrino mass scheme with
inverted mass hierarchy and $\sin^2\Theta_{13}\gsim 10^{-3}$.

We have introduced a simple algorithm
to identify the peaks in the presence of background fluctuations. This
method is based on the integration of the area around the expected
position of the peak.  By comparing the area distribution without and
with the spectral modulations induced by the Earth matter effects we
have analyzed the statistical significance of the result. As expected
 the presence of the core as well as a larger
difference in the initial spectra enhance the probability of
identifying the Earth effects.  
We have also presented a variation of the algorithm which improves its
efficiency significantly in the case of a water Cherenkov detector for
neutrino trajectories passing through the core.
Therefore we believe that more efficient algorithms could be developed.
The results we presented should be considered therefore as conservative
lower limits.

\section*{Acknowledgments}

We would like to thank Thomas Schwetz for useful discussions.  This
work was supported, in part, by the Deutsche Forschungsgemeinschaft
under grant No.\ SFB-375 and by the European Science Foundation (ESF)
under the Network Grant No.~86 Neutrino Astrophysics. M.K.\
acknowledges support by an Emmy-Noether grant of the Deutsche
Forschungsgemeinschaft, and R.T.\ a Marie-Curie-Fellowship of the
European Community.

\section*{References}

\end{document}